\documentclass[twocolumn,aps, showpacs, showkeys]{revtex4}
\usepackage{graphicx}% Include figure files
\begin{document}
%\draft

\title{Lagrangean description of nonlinear dust--ion acoustic
waves in dusty plasmas \footnote{Preprint; submitted to
{\textit{European Physical Journal D}.} }}

\author{I. Kourakis\footnote{On leave from: U.L.B. - Universit\'e Libre de Bruxelles,
Physique Statistique et Plasmas C. P. 231, Boulevard du Triomphe, B-1050 Brussels,
Belgium;
also:
Facult\'e des Sciences Apliqu\'ees - C.P. 165/81 Physique
G\'en\'erale, Avenue F. D. Roosevelt 49, B-1050 Brussels, Belgium;
\\Electronic address: \texttt{ioannis@tp4.rub.de}} and P. K.
Shukla\footnote{Electronic address: \texttt{ps@tp4.rub.de}}}
\affiliation{Institut f\"ur Theoretische Physik IV,
Fakult\"at f\"ur Physik und Astronomie, \\
Ruhr--Universit\"at Bochum, D-44780 Bochum, Germany}

\date{Submitted 23 March 2004}

\begin{abstract}
An analytical model is presented for the description of nonlinear
dust-ion-acoustic waves propagating in an unmagnetized, collisionless,
three component plasma composed of electrons, ions and inertial dust
grains. The formulation relies on a Lagrangean approach of the plasma
fluid model. The modulational  stability of the wave amplitude is
investigated. Different types of localized envelope electrostatic
excitations are shown to exist.
\end{abstract}

\pacs{52.27.Lw, 52.35.Fp, 52.35.Mw, 52.35.Sb}

\keywords{Electrostatic waves, ion--acoustic mode, nonlinear
waves.}

\maketitle

\section{Introduction}

In the last two decades, dusty plasmas (DP) have attracted a
great deal of attention due to a variety of new phenomena
observed in them and the novel physical mechanisms involved in
their description \cite{PSbook, Verheest}. In addition to known
plasma electrostatic modes \cite{Krall}, new oscillatory
modes arise in DP \cite{PSbook, Verheest}, among which the
\textit{dust-ion acoustic wave} (DIAW) and \textit
{dust acoustic waves} (DAW)  are of significant
interest in laboratory dusty plasma discharges.
In the DIAW  the restoring force comes from the
pressures of inertialess electrons, whereas the ion mass provides the
inertia, similar to the usual ion-acoustic waves in an electron-
ion plasma. Thus, the DIAW is characterized by a phase speed much smaller (larger)
than the ion (electron) thermal speed, and a frequency much higher than the
dust plasma frequency $\omega_{p, d}$; therefore, on the timescale of our
interest, stationary dust grains do not participate in the wave
dynamics; they just affect the equilibrium quasi-neutrality condition.
As a matter of fact, the DIAW phase velocity is higher than
that of IA waves, due to the electron density depletion in the
background plasma when dust grains are negatively charged;
quite remarkably, this fact results in suppression of the Landau damping
mechanism \cite{PSbook}, known to prevail over the IAW
propagation in an electron-ion plasma \cite{Krall}.

The linear properties of the IAWs have been quite extensively studied
and now appear well understood \cite{PSbook}. As far as nonlinear
effects are concerned, various studies have pointed out the possibility
of the formation of DIAW-related localized structures, due to a mutual
compensation between nonlinearity and dispersion, including small-amplitude
pulse solitons, shocks and vortices \cite{PSsolitons}.
Furthermore, the propagation of nonlinearly modulated DIA wave packets was
studied in Ref. \cite{IKPSDIAW}, in addition to the formation of
localized envelope soliton--modulated waves due to the modulational
instability of the carrier waves.  A very interesting known approach,
not yet included in our current knowledge with respect to the DIA plasma
waves, is the Lagrangean description of a nonlinear  wave profile.
In the context of electrostatic plasma waves, this formalism has
been employed in studies of electron plasma waves
\cite{Davidson1, Davidson2, Infeld} and, more recently,
ion-acoustic \cite{Chakra1} and dust-acoustic \cite{Chakra2}  waves.
Our aim here is to extend previous results by applying the Lagrangean
formalism to the description of nonlinear DIAWs propagating in dusty
plasmas.

We shall consider the nonlinear propagation of dust-ion-acoustic
waves in a collisionless plasma consisting of three distinct
particle species `$\alpha$': an inertial species of ions (denoted by
`$i$'; mass $m_i$, charge $q_i = + Z_i e$; $e$ denotes the absolute of the
electron charge), surrounded by an
environment of thermalized electrons (mass $m_e$, charge $- e$), and massive
dust grains (mass $M$, charge $q_d = s Z_d e$, both assumed constant for
simplicity); $Z_d$ denotes the charge state of dust grains; we leave the
choice of dust grain charge sign
$s = q_d/|q_d|$ ($= -1/+1$ for negative/positive dust charge)
open in the algebra. Charge neutrality is assumed at equilibrium.

\section{The model}

Let us consider the hydrodynamic--Poisson system of equations which describe the
evolution of the ion `fluid' in the plasma.  The ion number density $n_i$  is
governed by the continuity equation
\begin{equation}
\frac{\partial n_i}{\partial t} + \nabla  (n_i \,{\bf u}_i)= 0 \, ,
\label{densityequation}
\end{equation}
where the mean velocity ${\bf u}_i$ obeys
\begin{equation}
\frac{\partial {\bf u}_i}{\partial t} + {\bf u}_i  \cdot \nabla {\bf u}_i \,
= \, \frac{Z_i e}{m_i} {\bf E} \, = \, - \frac{Z_i e}{m_i}\,\nabla \,\Phi \, .
\end{equation}
The electric field ${\bf E} = -\nabla \,\Phi $
is related to the gradient of the wave potential $\Phi$, which is
obtained from Poisson's equation $\nabla \cdot {\bf E} = 4 \pi \, \sum q_s\, n_s $, viz.
\begin{equation}
\nabla^2 \Phi \,
%=\, - 4 \pi \, \sum q_s\, n_s \,
= \,4 \pi \,e
\,(n_c  \,+ n_h  \, - Z_i \,n_i)  \, . \label{Poisson}
\end{equation}
Alternatively, one may consider
\begin{equation}
\frac{\partial E}{\partial t} \, =\, - 4 \pi \, \sum_\alpha q_\alpha\, n_\alpha
\,u_\alpha  \, . \label{Poisson2}
\end{equation}
 We assume a near-Boltzmann distribution for the electrons,
 i.e. \(n_e \approx n_{e,0}\, \exp(e \Phi/k_B T_e) \,
\) ($T_e$ is the electron temperature and
$k_B$ is Boltzmann's constant). The
dust distribution is assumed stationary, i.e. $n_d \approx {\rm{const.}}$.
The overall quasi-neutrality condition at equilibrium then reads
\begin{equation}
Z_i \, n_{i, 0} \, + s Z_d \, n_{d} \,  -  n_{e, 0} \,  =\, 0 \, .
\label{neutrality}
\end{equation}

\subsection{Reduced Eulerian equations \label{reducedEuler}}

By choosing appropriate physical scales, Eqs. (1)-(3)
can be cast into a reduced (dimensionless) form. Let us
define the ion-acoustic speed $c_{s} = (k_B T_{e}/m_i)^{1/2}$.
An appropriate choice for the space and timescales,
$L$ and $T = L/c_{s}$, are the effective Debye length
$\lambda_{D} = (k_B T_{e}/4 \pi Z_i^2 n_{i, 0} e^2)^{1/2}
\equiv c_{s}/\omega_{p, i}$ and the  ion plasma period
$\omega_{p, i}^{-1} = (4 \pi n_{i, 0} Z_i^2 e^2/m_i)^{- 1/2}$,
respectively.  Alternatively, one might leave the choice of
$L$ (and thus $T = L/c_{s}$) arbitrary -- following
 an idea suggested in Refs. \cite{Chakra1, Chakra2}) -- which
 leads to the appearance of a dimensionless dispersion parameter
$\delta = 1/(\omega_{p, i} T) = \lambda_{D}/L$ in the formulae.
The specific choice of scale made above corresponds to $\delta = 1$
(implied everywhere in the following, unless otherwise stated);
however, we may keep the parameter $\delta$
 to `label' the dispersion term in the forthcoming formulae.

For one-dimensional wave propagation along the $x$ axis, Eqs.
(\ref{densityequation}) - (\ref{Poisson}) can now be written as
\begin{eqnarray}
\frac{\partial n}{\partial t} + \frac{\partial (n \, u)}{\partial x} & = & 0\, ,
\nonumber \\
\frac{\partial u}{\partial t} + u  \frac{\partial u}{\partial x} \, &
= & \,- \nabla \phi \,
,\nonumber \\
\delta^2 \, \frac{\partial^2 \phi}{\partial x^2} \, & = & \, \,(\hat n - n)\, ,
\label{reducedeqs}
\end{eqnarray}
where all quantities are dimensionless: $n = n_i/n_{i, 0}$,
$\mathbf{u} = \mathbf{u}_i/v_{0}$ and $\phi = \Phi/\Phi_0$; the
scaling quantities are, respectively: the equilibrium ion density
$n_{i, 0}$, the effective sound speed $v_{0} = c_{s}$
(defined above) and $\Phi_{0} = k_B T_{e}/(Z_i e)$.
The (reduced) electron and dust background density $\hat n$ is defined as
\begin{equation}
\hat n = \frac{n_{e}}{Z_i n_{i, 0}}\, e^{\phi/Z_i} + s
\frac{Z_d n_{d}}{Z_i n_{i, 0}}\, \equiv
\mu \, e^{\phi/Z_i} + 1 - \mu
\, , \label{defhatn1}
\end{equation}
where we have defined the DP parameter
$\mu = {n_{e, 0}}/({Z_i n_{i, 0}})$,
and made use of Eq. (\ref{neutrality}). Note
that both $n$ and $\hat n$ reduce to unity at equilibrium.

We shall define, for later reference,  the function $f(\phi) = \hat n$ -- given by Eq.
(\ref{defhatn1}) -- and its inverse function
\begin{equation}
f^{-1}(x) = Z_i \, \ln \biggl( 1 + \frac{x - 1}{\mu}
\biggr) \equiv g(x) \, , \label{defg} \end{equation}
viz. $f(\phi)
= x$ implies $\phi = f^{-1}(x) \equiv g(x)$.

We note that the dependence on the charge sign $s$ is now incorporated in
$\mu = 1 + s {Z_d n_{d, 0}}/({Z_i n_{i, 0}})$; retain that
$\mu < 1$ ($\mu > 1$) corresponds to negative (positive) dust.
Remarkably, since the dust-free
limit is recovered for $\mu = 1$, the results to be obtained in the
following are also straightforward valid for ion-acoustic waves
propagating in (dust-free) e-i plasma, upon setting $\mu = 1$ in the
formulae.

The well--known DIAW dispersion relation $\omega^2 = c_{s}^2
k^2/(k^2 \lambda_{D}^2 + 1)$ \cite{IKPSDIAW} is obtained from
Eqs. (\ref{densityequation}) to (\ref{neutrality}). On the other
hand, the system (\ref{reducedeqs}) yields the reduced relation
$\omega^2 = k^2/(\delta^2 k^2  + 1)$, which of course immediately
recovers the former dispersion relation upon restoring dimensions
(regardless, in fact, of one's choice of space scale $L$; cf.
definition of $\delta$). However, some extra qualitative information is
admittedly hidden in the
latter (dimensionless) relation. Should one consider a very long space
scale $L \gg \lambda_D$ (i.e. $\delta \ll 1$), one readily obtains $\omega
\sim k$ (unveiling the role of $\delta$ as a characteristic
dispersion control parameter). Finally, the opposite limit of
short $L$ (or infinite $\delta$) corresponds to ion plasma
oscillations (viz. $\omega = \omega_{p, i}$ = constant).

\subsection{Lagrangean description}

Let us introduce the Lagrangean variables $\{ \xi, \tau \}$, which
are related to the Eulerian ones $\{ x, t \}$ via
\begin{equation}
\xi \, = \, x\, - \int_0^\tau u(\xi ,\tau')\, d\tau' \, , \qquad
\qquad \tau = t \, . \label{Lagrange-def}
\end{equation}
See that they coincide at $t = 0$. Accordingly, the space and time
gradients are transformed as
\[
\partial/\partial x \rightarrow \alpha^{-1} \, \partial/\partial \xi \, ,
\qquad
\partial/\partial t \rightarrow \partial/\partial \tau -
\alpha^{-1} \, u \, \partial/\partial \xi\, , \] where we have
defined the quantity
\begin{equation}
\alpha(\xi, \tau) \equiv \frac{\partial x}{\partial \xi} = 1 +
\int_0^\tau d\tau' \frac{\partial }{\partial \xi} u(\xi ,\tau') \, .
\label{L0}
\end{equation}
Note that the convective derivative $D \equiv \partial/\partial
t + u \,\partial/\partial x$ is now plainly identified to $\partial/\partial
\tau$.  Also notice that $\alpha$ satisfies $\alpha(\xi, \tau = 0) = 0$ and
\begin{equation}
\frac{\partial \alpha(\xi, \tau)}{\partial \tau} =
\frac{\partial u(\xi, \tau)}{\partial \xi}
\label{property1}
\end{equation}
As a matter of fact, the Lagrangean transformation defined here reduces
to a Galilean transformation if one suppresses the evolution of $u$,
i.e. for $u = {\rm const.}$ (or $\partial u/\partial \tau =
\partial u/\partial \xi = 0$, hence $\alpha = 1$).
Furthermore, if one also suppresses the dependence in time $\tau$,
this transformation is reminiscent of the travelling wave ansatz
$f(x, t) = f(x - v t \equiv s)$, which is widely used in the Sagdeev
potential formalism \cite{PSsolitons, Sagdeev}.

The Lagrangean variable transformation defined above leads
to a new set of reduced equations
\begin{eqnarray}
n(\xi, \tau) &=& \alpha^{-1}(\xi, \tau) \, n(\xi, 0) \label{L1} \\
\frac{\partial u(\xi, \tau)}{\partial \tau} &=& \frac{Z_i e}{m_i}
E(\xi, \tau) \nonumber \\
&=& - \frac{Z_i e}{m_i} \,\alpha^{-1}(\xi, \tau) \,\frac{\partial
\phi(\xi, \tau)}{\partial \xi}  \qquad \label{L2} \\
\alpha^{-1}(\xi, \tau) \,\frac{\partial E(\xi, \tau)}{\partial
\xi} &=& 4 \pi
Z_i e [n(\xi, \tau) - \hat n \,n_{i, 0}] \label{L3} \\
\biggl( \frac{\partial }{\partial \tau} - \alpha^{-1} u \,
\frac{\partial }{\partial \xi} \biggr) E(\xi, \tau) &=& - 4 \pi
Z_i e n(\xi, \tau) u(\xi, \tau) \, , \label{L4}
\end{eqnarray}
where we have temporarily restored dimensions for physical
transparency; recall that the (dimensionless) quantity $\hat n $,
which is in fact a function of $\phi$, is given by (\ref{defhatn1}).
One immediately recognizes the role of the (inverse of the) function
$\alpha(\xi, \tau)$ as a density time evolution operator; cf. Eq.
(\ref{L1}) \cite{comment1}. Poisson's equation is now obtained by eliminating
$\phi$ from Eqs. (\ref{L2}, \ref{L3})
\begin{equation}
\alpha^{-1} \,\frac{\partial }{\partial \xi} \biggl( \alpha^{-1}
\, \frac{\partial \phi}{\partial \xi} \biggr) \,= - 4 \pi Z_i e (n
- \hat n \,n_{i, 0}) \, . \label{L5}
\end{equation}
Note that a factor $\delta^2$ should appear in the left-hand side if
one rescaled Eq. (\ref{L5}) as described above; cf. the last of
Eqs. (\ref{reducedeqs}). This will be retained for later
reference, with respect to the treatment suggested in Ref.
\cite{Chakra1} (see discussion below).

 In principle, our aim is to solve the system of Eqs.
(\ref{L1}) to (\ref{L4}) or, by eliminating $\phi$, Eqs. (\ref{L1}),
(\ref{L2}) and (\ref{L5}) for a given initial condition $n(\xi, \tau=0)
= n_0(\xi)$, and then make use of the definition
(\ref{Lagrange-def}) in order to invert back to the Eulerian
arguments of the state moment variables (i.e. density, velocity
etc.). However, this abstract scheme is definitely
not a trivial  task to accomplish.

\section{Nonlinear dust-ion acoustic oscillations}

Multiplying Eq. (\ref{L3}) by $u(\xi, \tau)$ and then adding to
Eq. (\ref{L4}), one obtains
\begin{equation}
\frac{\partial E(\xi, \tau)}{\partial \tau} = - 4 \pi Z_i e n_{i,
0}\, \hat n \, u(\xi, \tau) \label{L6} \, .
\end{equation}
Combining with Eq. (\ref{L2}), one obtains
\begin{equation}
\frac{\partial^2 u}{\partial \tau^2} = - \omega_{p, i}^2 \, \hat n
\, u \, ,\label{NLoscil}
\end{equation}
where $\omega_{p, i}$ is the ion plasma frequency (defined above).
Despite its apparent simplicity, Eq. (\ref{NLoscil}) is
{\em{neither}} an ordinary differential equation (ODE) -- since
all variables depend on {\em{both}} time $\tau$ and space $\xi$ --
{\em{nor}} a closed evolution equation for the mean velocity
$u(\xi, \tau)$: note that the (normalized) background particle
density $\hat n$ depends on the potential $\phi$ and on the plasma
parameters; see its definition (\ref{defhatn1}). The evolution of
the potential $\phi(\xi, \tau)$, in turn, involves $u(\xi, \tau)$
(via the quantity $\alpha(\xi, \tau)$) and the ion density $n(\xi,
\tau)$.

Eq. (\ref{NLoscil}) suggests that the system performs nonlinear
oscillations at a frequency $\omega = \omega_{p, i} \, {\hat n}^{1/2}$.
Near equilibrium, the quantity ${\hat n}$ is approximately equal to unity and
one plainly recovers a linear oscillation at the ion plasma frequency $\omega_{p, i}$.
Quite  unfortunately this apparent simplicity, which might
in principle enable one to solve for $u(\xi, \tau)$ and then obtain
$\{ \xi, \tau \}$ in terms of $\{ x, t \}$ and vice versa (cf.
Davidson's treatment for electron plasma oscillations in Ref.
\cite{Davidson2}; also compare to Ref. \cite{Infeld}, setting
$\gamma = 0$ therein), is absent in the general (off-equilibrium)
case where the plasma oscillations described by Eq. (\ref{NLoscil})
are intrinsically {\em{nonlinear}}.

Since Eq. (\ref{NLoscil}) is in general not a closed equation for
$u$, unless the background density $\hat n$ is constant (i.e.
independent of $\phi$, as in Refs.  \cite{Davidson2, Infeld}), one
can \emph{neither} apply standard methods involved in the
description of nonlinear oscillators on Eq. (\ref{NLoscil}) (cf.
Ref. \cite{Infeld}), \emph{nor} reduce the description to a study
of Eqs. (\ref{NLoscil}, \ref{L6}) (cf. Ref. \cite{Davidson1}), but
rather has to retain all (or rather five) of the evolution
equations derived above, since five inter-dependent dynamical
state variables (i.e. $n$, $u$, $E$, $\phi$ and $\alpha$) are
involved. This procedure will be exposed in the following Section.

\section{Perturbative nonlinear Lagrangean treatment}

Let us consider weakly nonlinear oscillations performed by our system
close to (but not at) equilibrium.
The basis of our study will be the reduced system of equations
\begin{eqnarray}
\frac{\partial }{\partial \tau} (\alpha \, n) = 0 \, , \nonumber \\
\frac{\partial u}{\partial \tau} = E \, , \nonumber \\
\frac{\partial E}{\partial \xi} = (n - \hat n)\, \alpha \, , \nonumber \\
\alpha \, E = - \frac{\partial  \phi}{\partial \xi} \, , \nonumber \\
\frac{\partial \alpha}{\partial \tau} =
\frac{\partial u}{\partial \xi} \, , \label{system-reduced}
\end{eqnarray}
which follow from the Lagrangean Eqs. (\ref{L1}) to (\ref{L5}) by
scaling over appropriate quantities, as described in \S
\ref{reducedEuler} \cite{comment2}. This system describes the
evolution of the state
 vector, say ${\mathbf{S}} = (\alpha, n, u, E, \phi)$ ($\in \Re^5$), in the
 Lagrangean coordinates defined above. We will consider small deviations
 from the equilibrium state ${\mathbf{S}_0} = (1, 1, 0, 0, 0)^T$, by
 taking ${\mathbf{S}} = {\mathbf{S}}^{(0)} + \epsilon {\mathbf{S}_1}^{(0)} +
 \epsilon^2 {\mathbf{S}_2}^{(0)} + ...$, where $\epsilon$ ($\ll 1$) is a
 smallness parameter.  Accordingly, we shall Taylor develop the quantity
 $\hat n(\phi)$ near $\phi \approx 0$, viz. $\phi \approx \epsilon \phi_1 +
 \epsilon^2 \phi_2 + ...$, in order to express $\hat n$ as
 \begin{eqnarray}
 \hat n & \approx & 1 + c_1 \phi + c_2 \phi^2 + c_3 \phi^3 + ...
 \nonumber \\
 & = &  1 + \epsilon c_1 \phi_1 + \epsilon^2
 (c_1 \phi_2 + c_2 \phi_1^2) \nonumber \\
 &  &  + \epsilon^3
 (c_1 \phi_3 + 2 c_2 \phi_1 \phi_2 + c_3 \phi_1^3) + ... \, ,
 \end{eqnarray}
 where the coefficients $c_j$ ($j=1, 2, ...$), which are determined from
 the definition (\ref{defhatn1}) of $\hat n$,
 contain all the essential dependence on the plasma parameters,
 e.g. $\mu$;
 making use of $e^{x} \approx \sum_{n=0}^\infty {x}^n/n!$,
 one readily obtains
 \[
 c_1 = \mu/Z_i \, , \qquad c_2 = \mu/(2 Z_i^2)  \, , \qquad c_2 = \mu/(6 Z_i^3)
 \, . \]
Remember that for $\mu = 1$ (i.e.  for vanishing dust) one recovers the
expressions for IAWs in e-i plasma.

Following the standard reductive perturbation technique
\cite{redpert}, we shall consider the stretched (slow) Lagrangean
coordinates \( Z \,= \, \epsilon (\xi - \lambda \,\tau) \, , \quad
T \,= \, \epsilon^2 \, \tau\) (where $\lambda \in \Re$ will be
determined later). The perturbed state of (the $j-$th --- $j = 1,
..., 5$ --- component of) the state vector  ${\mathbf{S}}^{(n)}$ is
assumed to depend on the fast scales via the carrier phase $\theta
= k \xi - \omega \tau$, while the slow scales enter the argument
of the ($j-$th element's) $l-$th harmonic amplitude $S_{j,
l}^{(n)}$, viz. \( S{j}^{(n)} \,= \, \sum_{l=-\infty}^\infty
\,S_{j, l}^{(n)}(Z, \, T)
 \, e^{i l (k \xi - \omega \tau)}\) (where $S_{j,-l}^{(n)} =
 {S_{j, l}^{(n)}}^*$ ensures reality).
Treating the derivative operators as
\[
\frac{\partial}{\partial \tau} \rightarrow \frac{\partial}{\partial
\tau} - \epsilon \, \lambda \, \frac{\partial}{\partial Z} +
\epsilon^2 \, \frac{\partial}{\partial T} \, , \qquad
\frac{\partial}{\partial \xi}
\rightarrow \frac{\partial}{\partial \xi} + \epsilon \,
\frac{\partial}{\partial Z} \, ,
\]
and substituting into the system of evolution equations, one
obtains an infinite series in both (perturbation order)
$\epsilon^n$ and (phase harmonic) $l$. The standard perturbation
procedure now consists in solving in successive orders $\sim
\epsilon^n$ and substituting in subsequent orders. The method involves
a tedious calculation which is
however  straightforward; the details of the method
are presented e.g. in Ref. \cite{IKPSDIAW},
so only the essential stepstones need to be provided here.

The equations obtained for $n = l = 1$ determine the first harmonics
of the perturbation
\begin{eqnarray}
n_1^{(1)} \, = - \alpha_1^{(1)} \, =
\, ({k^2}/{\omega^2})  \psi \, , \nonumber
\\
u_1^{(1)} \, = (k/\omega) \psi
\, , \qquad
E_1^{(1)} \, = - i k \psi
\label{1st-order-corrections}
\end{eqnarray}
where $\psi$ denotes the potential correction $\phi_1^{(1)}$.
The cyclic frequency $\omega$ obeys the dispersion relation
\( \omega^2\,  = {k^2}/({k^2 + s c_1})\),
which exactly recovers, once dimensions are restored, the standard
IAW dispersion relation \cite{Krall} mentioned above.

Proceeding in the same manner, we obtain the second order
quantities, namely the amplitudes of the second harmonics
$\mathbf{S}_2^{(2)}$ and constant (`direct current') terms
$\mathbf{S}_0^{(2)}$, as well as a finite contribution
$\mathbf{S}_1^{(2)}$ to the first harmonics; as expected from
similar studies, these three (sets of 5, at each $n, l$)
quantities are found to be proportional to $\psi^2$, $|\psi|^2$
and $\partial \psi/\partial Z$ respectively; the lengthy
expressions are omitted here for brevity. The ($n = 2$, $l=1$)
equations provide the compatibility condition: \( \lambda \,=
\omega (1 - \omega^2)/k = {d \omega}/{d k}\); $\lambda$ is
therefore the group velocity $v_g(k) = \omega'(k)$ at which the
wave envelope propagates. It turns out that $v_g$ decreases with
increasing wave number $k$; nevertheless, it always remains
positive.

In order $\sim \epsilon^3$, the equations for $l = 1$
yield an explicit compatibility condition in the form of a
nonlinear Schr\"odinger--type equation (NLSE)
\begin{equation}
i\, \frac{\partial \psi}{\partial T} + P\, \frac{\partial^2
\psi}{\partial Z^2} + Q \, |\psi|^2\,\psi = 0 \, .  \label{NLSE}
\end{equation}
Recall that $\psi\, \equiv \,  \phi_1^{(1)}$ denotes the amplitude of
the first-order electric potential perturbation.
The `slow' variables $\{ Z, T \}$ were defined above.

The {\em dispersion coefficient} $P$ is related to the curvature
of the dispersion curve as \( P \,  = \, \omega''(k)/{2} \,= - 3
\omega^3 (1- \omega^2)/(2 k^2)\). One may easily check that $P$ is
negative (for all values of $k$).

 The {\em nonlinearity coefficient} $Q$ is due to carrier
wave self-interaction. It is given by the expression
\begin{equation}
Q = + \frac{\omega^3}{12 \, k^4}
\frac{\mu}{Z_i^4}\, \biggl[ 3 Z_i^3 k^6
- 3 (\mu + 4) Z_i^2 k^4
%\qquad \nonumber \\ \qquad
+ 3 (1 - 2 \mu - 5 \mu^2) Z_i k^2 - \mu (3 \mu - 1)^2
\biggr] \, .
 \label{Qcoeff}
\end{equation}
where the coefficients $c_{1, 2, 3}$ were defined above.

For low wavenumber $k$, $Q$ goes to $- \infty$ as
 \[
Q \approx - \frac{(3 \mu - 1)^2 \mu^{1/2}}{12 \, Z_i^{5/2}} \, \frac{1}{k} \, .
\]

\subsection{Modulational stability analysis}

According to the standard analysis \cite{Hasegawa}, we can
linearize around the plane wave solution of the NLSE (\ref{NLSE})
\(\psi \, = \, {\hat \psi} \, e^{i Q |\hat \psi|^2 \tau} \, + \,
c.c. \, , \) ($c.c.$: complex conjugate) -- notice the
amplitude dependence of the frequency shift $\Delta \omega =
\epsilon^2 Q |\hat \psi|^2$ -- by setting \({\hat \psi} \, = \,
{\hat \psi}_0 \, + \, \epsilon \, {\hat \psi}_1 \, , \) and then
assuming the perturbation ${\hat \psi}_1$ to be of the form:
${\hat \psi}_1 \, = \, {\hat \psi}_{1, 0} \,e^{i ({\hat k} \zeta -
{\hat \omega} \tau)} \, + \, c.c.$. Substituting into
(\ref{NLSE}), one thus readily obtains \( \hat \omega^2 \, = \,
P^2 \, \hat k^2 \, \biggl(\hat k^2 \, - \, 2 ({Q}/{P}) |\hat
\psi_{1, 0}|^2 \biggr) \). The wave will thus be {\em stable}
($\forall \, \hat k$) if the product $P Q$ is negative. However,
for positive $P  Q > 0$, instability sets in for wavenumbers
below a critical value $\hat k_{cr} = \sqrt{2 Q/P}\, |\hat
\psi_{1, 0}|$, i.e. for wavelengths above a threshold
$\lambda_{cr} = 2 \pi/\hat k_{cr}$; defining the instability
growth rate \( \sigma = |Im\hat\omega(\hat k)| \), we see that it
reaches its maximum value for $\hat k = \hat k_{cr}/\sqrt{2}$,
viz.
\[ \sigma_{max} =
|Im\hat\omega|_{\hat k = \hat k_{cr}/\sqrt{2}} \,=\, | Q |\, |\hat
\psi_{1, 0}|^2  \, .
\]
We see that the instability condition depends only on
the sign of the product $P Q$, which may be studied numerically,
relying on the expressions derived above.

\subsection{Finite amplitude nonlinear excitations}

The NLSE (\ref{NLSE}) is long known to possess distinct types of
localized constant profile (solitary wave) solutions, depending on
the sign of the product $P Q$ \cite{Hasegawa, Fedele, IKPSDIAW}.
Remember that this equation here describes the evolution of the
wave's envelope, so these solutions represent slowly varying
localized envelope structures, confining the (fast) carrier wave.
The analytic form of these excitation can be found in the literature
(see e.g. in \cite{IKPSDIAW} for a brief review) and need not be derived
here in detail. Let us however briefly summarize those results.

Following Ref. \cite{Fedele}, we may seek a solution of Eq.
(\ref{NLSE}) in the form \( \psi(\zeta, \tau) = \rho(Z, T) \,
e^{i\,\Theta(\zeta, \tau) } + {\rm c.c.}\),  where $\rho$,
$\sigma$ are real variables which are determined by substituting
into the NLSE and separating real and imaginary parts. The
different types of solution thus obtained are summarized in the
following.

\begin{figure}[htb]
 \centering
 \resizebox{3in}{!}{
\includegraphics{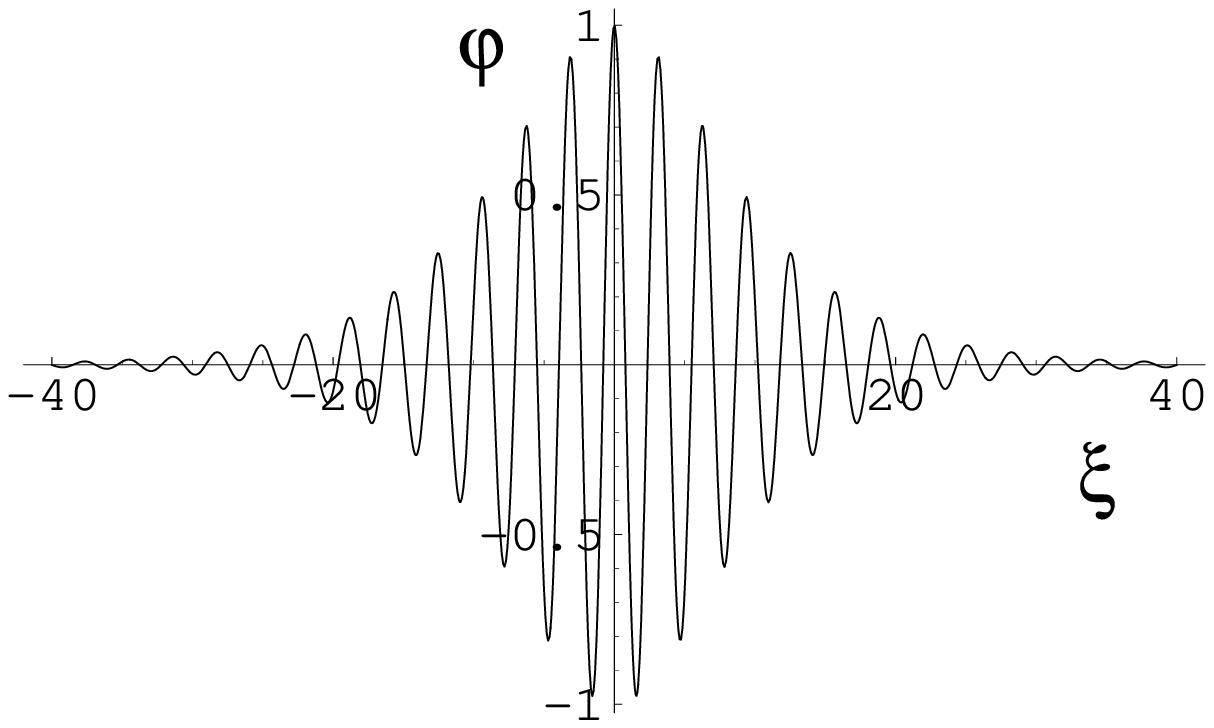}
} \vskip 1 cm
\resizebox{3in}{!}{
\includegraphics{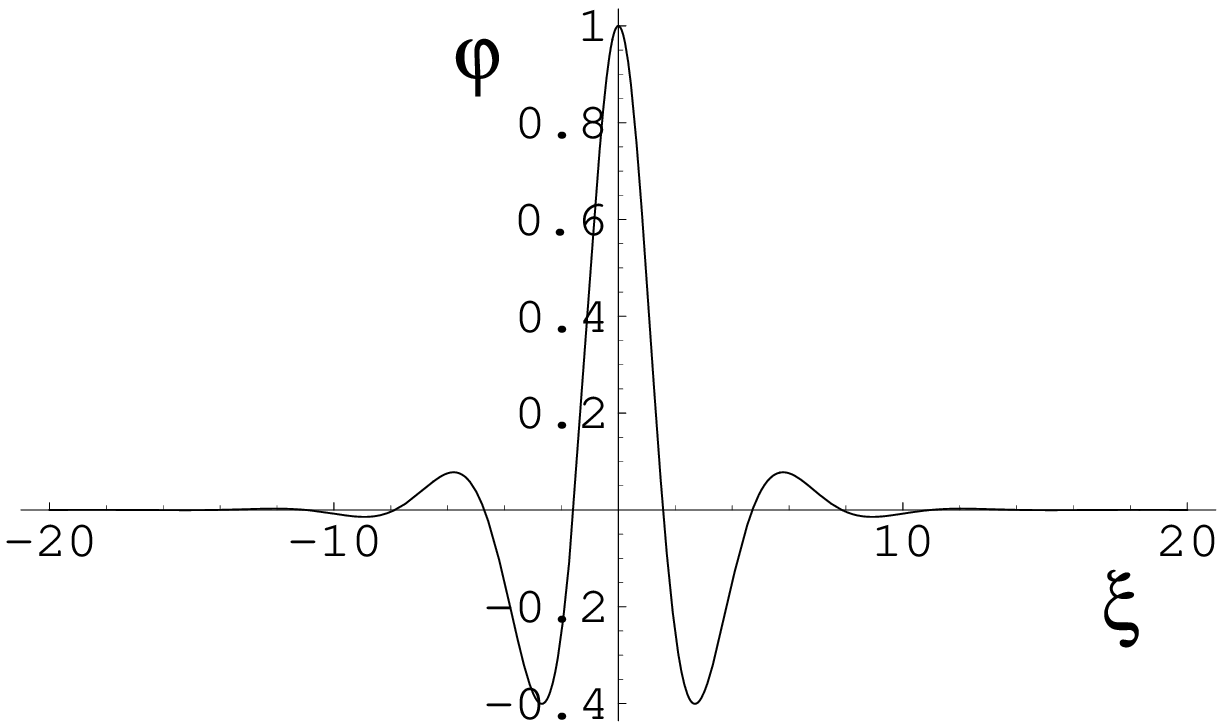}
} \vskip 1 cm
 \resizebox{2.8in}{!}{
 \includegraphics[]{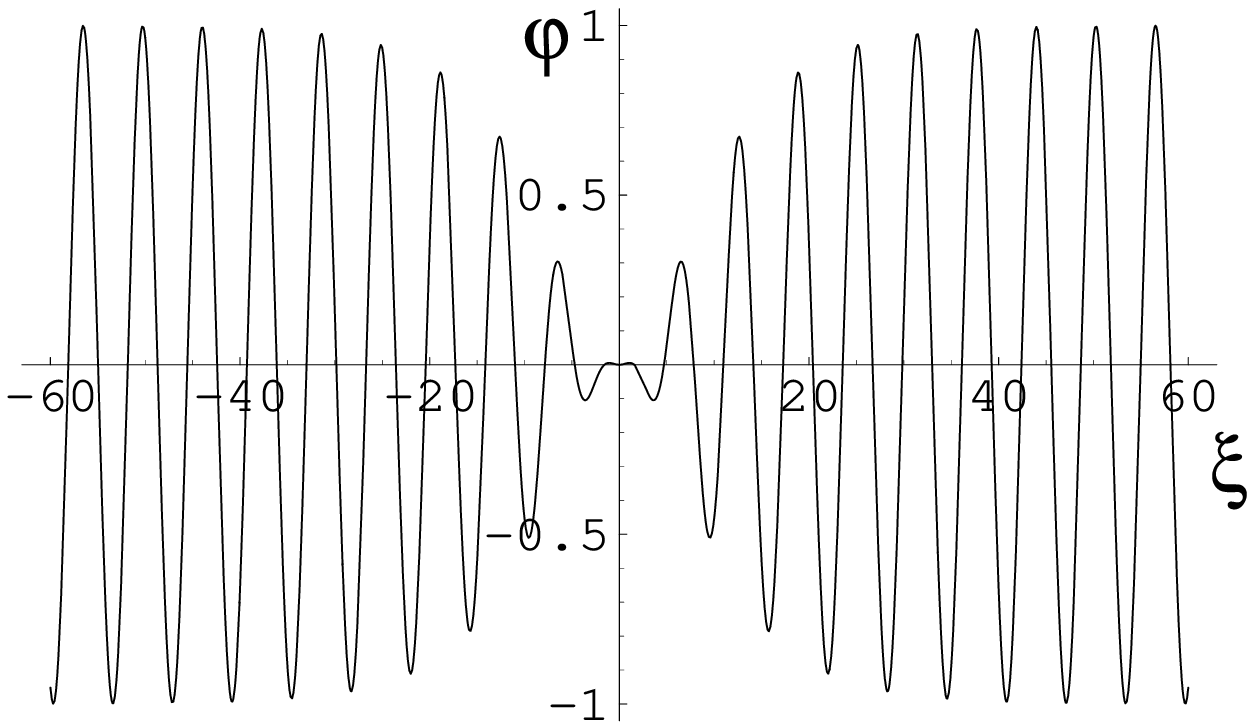}}
\\
\vskip 1 cm \resizebox{2.8in}{!}{
\includegraphics{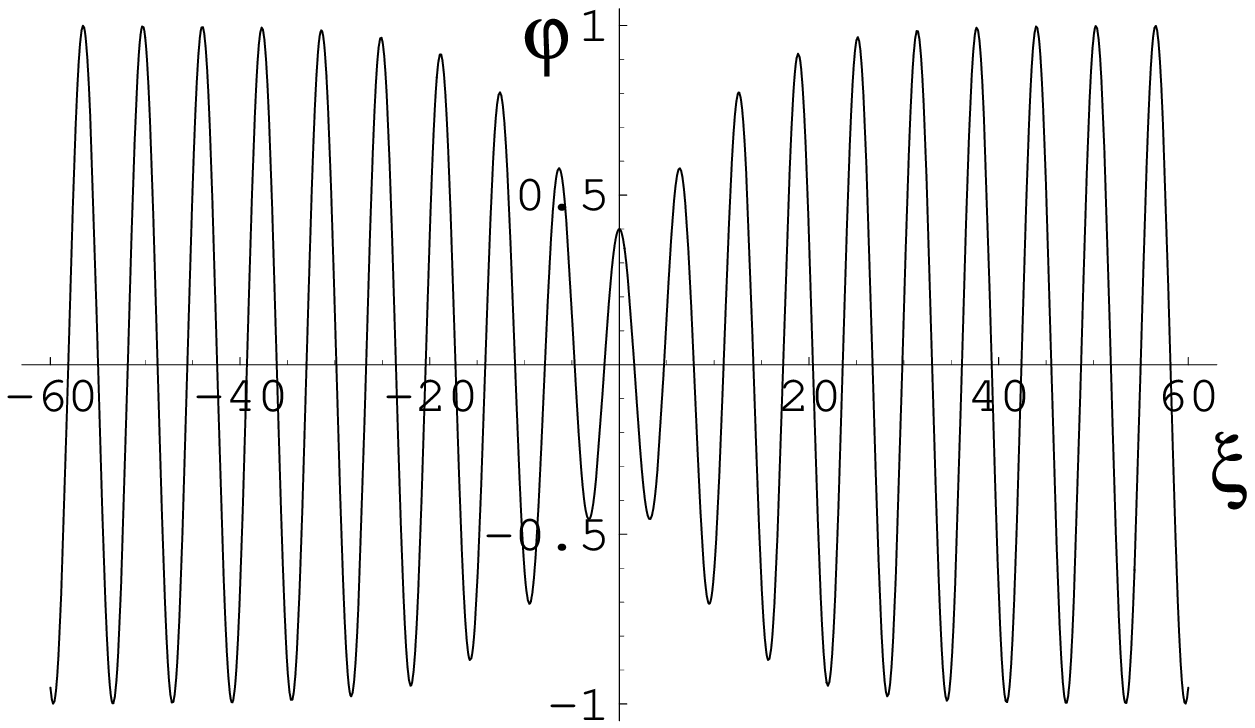}
} \caption{} \label{figure1}
\end{figure}

For $P Q > 0$ we find the {\em (bright) envelope soliton}
\begin{equation}
\rho = \pm \rho_0 \, sech \biggl(\frac{Z - u_e\, \tau}{L} \biggr)
\, , \quad  \Theta = \frac{1}{2 P} \, \bigl[ u_e Z - (\Omega +
\frac{1}{2} u_e^2) T \bigr] \, , \label{bright}
\end{equation}
which represents a localized pulse travelling at the envelope
speed $u_e$ and oscillating at a frequency $\Omega$ (at rest). The
pulse width $L$ depends on the maximum amplitude square $\rho_0$
as \( L = ({2 P}/{Q })^{1/2}/\rho_0 \). Since the product $P Q$ is
always positive for long wavelengths, as we saw above, this type
of excitation will be rather privileged in dusty plasmas. The bright-type
envelope soliton is depicted in Fig. \ref{figure1}a, b.

For $P Q < 0$, we obtain the {\em dark} envelope soliton ({\em hole})
\cite{Fedele}
\begin{eqnarray}
\rho & = & \pm \rho_1 \, \biggl[ 1 -  sech^2 \biggl(\frac{Z - u_e
T}{L'} \biggr)\biggr]^{1/2}  \nonumber \\
&=& \pm \rho_1 \,
 \tanh \biggl(\frac{Z - u_e\, T}{L'}
\biggr) \, ,
\nonumber \\
\Theta & = & \frac{1}{2 P} \, \biggl[ u_e Z \, -
\biggl(\frac{1}{2} u_e^2 - 2 P Q \rho_1 \biggr) \,T \biggr] \, ,
\label{darksoliton}
\end{eqnarray}
which represents a localized region of negative wave density
(shock) travelling at a speed $u_e$; see Fig. \ref{figure1}c.
Again, the pulse width
depends on the maximum amplitude square $\rho_1$ via \( L' = (2
\bigl|{P}/{Q}\bigr|)^{1/2}/\rho_1 \).

Finally, still for $P Q < 0$, one also obtains the {\em gray} envelope
solitary wave \cite{Fedele}
\begin{equation}
\rho = \pm \rho_2 \, \biggl[ 1 - a^2\, sech^2 \biggl(\frac{Z - u_e
T}{L''} \biggr)\biggr]^{1/2} \, , \label{greysoliton}
\end{equation}
which also represents a localized region of negative wave density.
Comparing to the dark soliton (\ref{darksoliton}), we note that the
maximum amplitude $\rho_2$ is now finite (non-zero) everywhere;
see Fig. \ref{figure1}d. The
 the pulse width of this gray-type excitation
\( L'' = \sqrt{2 | {P}/{Q} |}/(a \,\rho_2) \) now also depends on
an independent parameter $a$ which represents the modulation depth
($0 < a \le 1$). The lengthy expressions which determine the phase
shift $\Theta$ and the parameter $a$, which are omitted here for
brevity, can be found in Refs. \cite{Fedele, IKPSDIAW}. For $a = 1$,
one recovers the {\em dark} soliton presented above.

An important qualitative result to be retained is that the
envelope soliton width $L$ and maximum amplitude $\rho$ satisfy $L
\rho \sim \sqrt{P/Q}$ (see above), and thus depend on (the ratio
of) the coefficients $P$ and $Q$; for instance, regions with
higher values of $P$ (or lower values of $Q$) will support wider
(spatially more extended) localized excitations, for a given value
of the maximum amplitude. Contrary to the KdV soliton picture, the
width of these excitations does not depend on their velocity. It
does, however, depend on the plasma parameters, e.g. here $\mu$.

\begin{figure}[htb]
 \centering
 \resizebox{4in}{!}{
\includegraphics{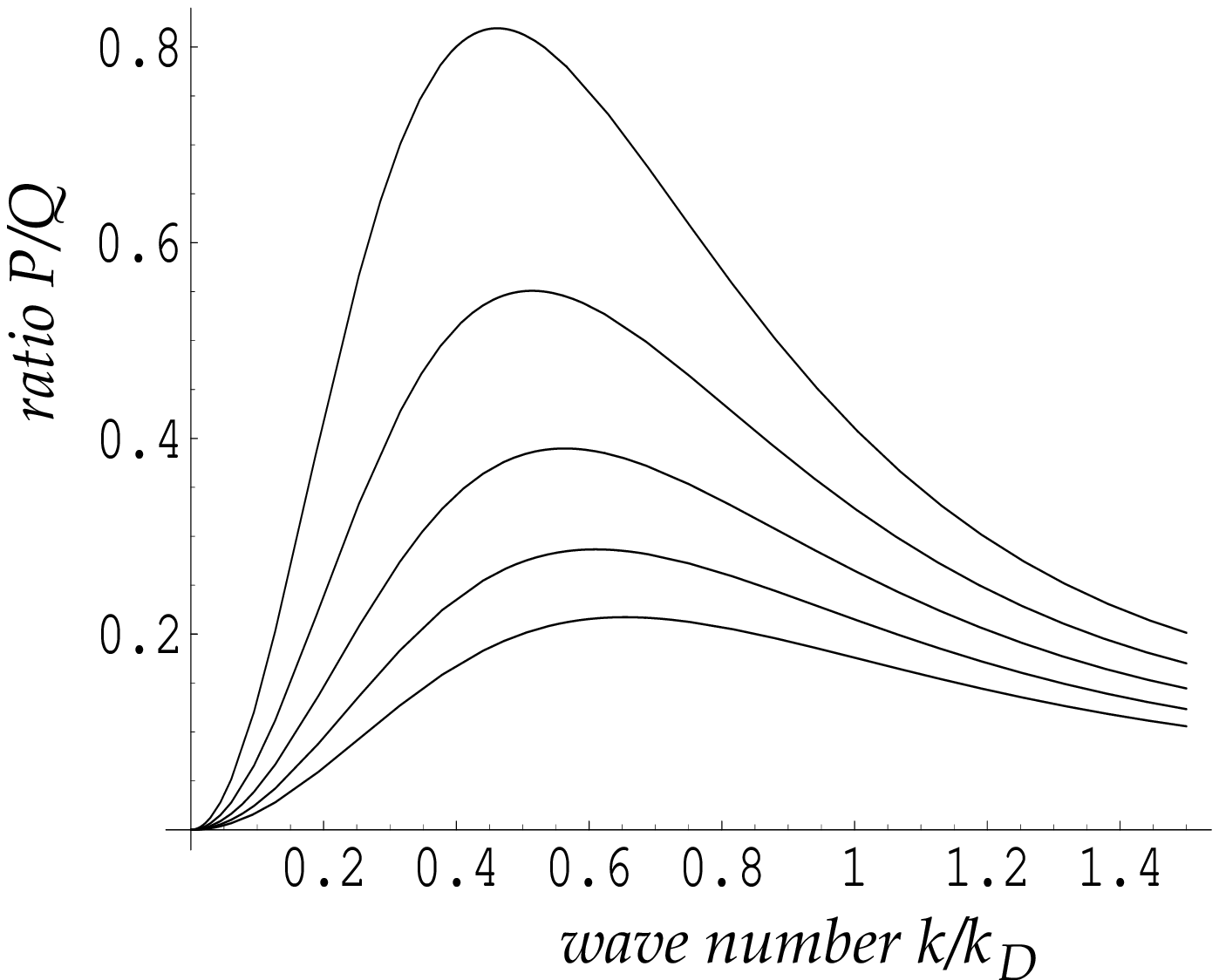}
} \caption{} \label{figure2}
\end{figure}

The localized envelope excitations presented above represent the slowly varying
envelope which confines the (fast) carrier space and time oscillations, viz.
$\phi = \Psi(X, Z) \cos(k \xi - \omega \tau)$ for the electric potential
$\phi$ (and analogous expressions for the density $n_i$ etc.; cf.
(\ref{1st-order-corrections})).
The qualitative characteristics (width, amplitude) of these excitations,
may be investigated
by a numerical study of the ratio $P/Q \equiv \eta(k; \mu)$: recall that
its sign determines the type (bright or dark) of the excitation,
while its (absolute) value determines its width for a given
amplitude (and vice versa). In Fig. \ref{figure2}
we have depicted the behaviour of $\eta$ as a
function of the wavenumber $k$ and the parameter
$\mu$: higher values of $\mu$ correspond to lower curves.
Remember that, for any  given wavenumber $k$, the dust concentration
(expressed via the value of $\mu$) determines the soliton width $L$
(for a given amplitude $\rho$; see discussion above) since
$L \sim \eta^{1/2}/\rho$. Therefore, we see that the addition of
{\em{negative}}  dust generally ($\mu < 1$) results to higher values of  $\eta$
(i.e. wider or higher solitons), while
{\em{positive}} dust ($\mu > 1$) has the opposite effect:
it reduces the value of  $\eta$ (leading to narrower or shorter solitons).
In a rather general manner, bright type solitons (pulses) seem to be rather
privileged, since the ratio $\eta$ (or the product $P Q$) of the coefficients
$P$ and $Q$ is positive in most of the $k, \mu$ plane of values.
One exception seems to be very the region of {\em{very}}
low values of $\mu$ (typically below $0.2$), which develops a negative tail
of $\eta$ for small $k$ ($< 0.3 \, k_D$):
thus, a very high ($> 80$ per cent) electron depletion results in
pulse destabilization in favour of dark-type excitations
(Fig. \ref{figure1}c, d).
Strictly speaking, $\eta$ also becomes negative for very high wave number values
($> 2.5 \, k_D$);
nevertheless, we neglect -- for rigor -- this region from the analysis,
in this (long wavelength $\lambda$)  fluid picture
(for a weak dust presence, short $\lambda$ DIAWs may be quite
strongly damped; however, this result may still be interesting
for a strong presence of dust, when Landau damping is not a significant issue
\cite{PSbook}).

\section{Relation to previous works:
an approximate nonlinear Lagrangean treatment}

By combining the Lagrangean system of Eqs. (\ref{L1}) to (\ref{L5}),
one obtains the (reduced) evolution equation
\begin{equation}
\frac{\partial^2}{\partial \tau^2} \biggl( \frac{1}{n}  \biggr) =
- \frac{1}{n_0} \frac{\partial }{\partial \xi} \, \biggl[
\frac{n}{n_0} \,\frac{\partial }{\partial \xi} \, g(w) \biggr] \,
, \label{Lfinal}
\end{equation}
where the function $g(x)$ [defined in Eq. (\ref{defg})] is
evaluated at
\[w(n) = n \biggl[1 - \delta^2 \frac{\partial^2}{\partial \tau^2}
\biggl( \frac{1}{n} \biggr) \biggr] \, .\] Note that the ion
density $n$ has been scaled by its equilibrium value $n_{i, 0}$,
to be distinguished from the initial condition $n_0 = n(\xi,
\tau = 0)$.

Despite its complex form, the
nonlinear evolution equation (\ref{Lfinal}) can be
solved exactly by considering different special cases, as
regards the order of magnitude of the dispersion--related
parameter $\delta$. This treatment,
based on Ref. \cite{Chakra1},
will only be briefly summarized here, for the sake of reference.

%\subsection{Linear oscillations}

First, one may consider very short scale variations, i.e. $L \ll \lambda_D$
(or $\delta \gg 1$). This amounts to neglecting collective effects, so
oscillatory motion within a Debye sphere is essentially decoupled from
neighboring ones. By considering \(w(n) \approx - \delta^2 \, n \,
{\partial^2}\bigl( {1}/{n} \bigr)/{\partial \tau^2} \)
and $\phi \approx 0$ (i.e. $\hat n \approx 1$),
one may combine Eqs. (\ref{L4}) and
(\ref{Lfinal}) into
\begin{equation}
\biggl( \frac{\partial^2 }{\partial \tau^2} + \omega_{p, i}^2 \biggr)
\biggl( \frac{1}{n} - 1 \biggr) = 0 \, ,
\end{equation}
which, imposing the initial condition $n(\xi, 0) = n_0(\xi)$, yields the
solution
\begin{equation}
n(\xi, \tau) = \frac{n_0(\xi)}{\frac{n_0(\xi)}{n_{i, 0}} +
\bigl(1 - \frac{n_0(\xi)}{n_{i, 0}} \bigr) \cos \omega_{p, i}\tau} \, .
\end{equation}
Note that if the system is initially at equilibrium,
viz. $n_0(\xi) = n_{i, 0}$, then it remains so at all times $\tau > 0$.
Now, one may go back to Eq. (\ref{L1}) and solve for $\alpha(\xi, \tau)$,
which in turn immediately provides the mean fluid velocity $u$
\[
u(\xi, \tau) = \omega_{p, i} \sin \omega_{p, i}  \tau \int_{\xi_0}^\xi
\biggl(1 - \frac{n_0(\xi')}{n_{i, 0}} \biggr) d\xi'
\]
via (\ref{L0}), and then
$E(\xi, \tau)$ and $\phi(\xi, \tau)$. Finally, the variable
transformation (\ref{Lagrange-def}) may now be inverted, immediately providing
the Eulerian position $x$ in terms of $\xi$ and $\tau$. We shall not go into
further details regarding this procedure, which is essentially analogue (yet not
identical) to Davidson's treatment of electron plasma oscillations.

%\subsection{Nonlinear oscillations}

Quite interestingly, upon neglecting the dispersive effects, i.e. setting
$\delta = 0$, Eq. (\ref{Lfinal}) may be solved by separation of variables,
and thus shown to possess a nonlinear special solution in the form of
a product, say $n(\xi,\tau) = n_1(\xi) n_2(\tau)$ \cite{comment2}.
This calculation was put forward in Ref. \cite{Chakra1}
(where the study of IAW -- in a single electron temperature plasma --
was argued to rely on an equation quasi-identical to Eq. (\ref{Lfinal})).
However, the solution thus obtained relies on doubtful physical grounds,
since the assumption $\delta \approx 0$, which amounts to remaining close to
equilibrium -- cf. the last of Eqs. (\ref{reducedeqs}),
implies an infinite space scale $L$ (recall the definition of $\delta$),
contrary to the very nature of the (localized) nonlinear excitation itself.
Rather not surprisingly, this solution was shown in Ref. \cite{Chakra1}
to decay fast in time, in both Eulerian and Lagrangean coordinates.
Therefore, we shall not pursue this analysis any further.

\section{Discussion and conclusions}

We have studied the nonlinear propagation of dust ion acoustic waves propagating
in a dusty plasma.  By employing a Lagrangean formalism, we have investigated the
modulational stability of the amplitude of the propagating dust ion acoustic
oscillations and have shown that these electrostatic waves may become unstable,
due to self interaction of the carrier wave.  This instability may either lead
to wave collapse or to wave energy localization, in the form of propagating localized
envelope structures.  We have provided an exact set of analytical expressions for these
localized excitations.

This study complements similar investigations which relied
on an Eulerian formulation of the dusty plasma fluid model \cite{IKPSDIAW}.
In fact, the Lagrangean picture provides a strongly modified nonlinear
stability profile for the wave amplitude, with respect to the previous
(Eulerian) description; this was intuitively expected, since the passing
to Lagrangean variables involves an inherently nonlinear transformation,
which inevitably modifies the nonlinear evolution profile of the system
described. However, the general qualitative result remains in tact:
the dust ion acoustic-type electrostatic plasma waves may propagate in the form of
localized envelope excitations, which are formed as a result of the
mutual balance between dispersion and nonlinearity in the plasma fluid.
More sophisticated descriptions, incorporating e.g. thermal or
collisional effects, may be elaborated in order to refine
the parameter range of the problem, and may be reported later.

\medskip

\begin{acknowledgments}
This work was supported by the European Commission (Brussels)
through the Human Potential Research and Training Network via the
project entitled: ``Complex Plasmas: The Science of Laboratory
Colloidal Plasmas and Mesospheric Charged Aerosols'' (Contract No.
HPRN-CT-2000-00140).
\end{acknowledgments}

\newpage

% FIGURE CAPTIONS

\centerline{\textbf{Figure Captions}}

\vskip 1cm

Figure 1.

A heuristic representation of wave packets modulated by
solutions of the NLS equation. These envelope excitations are of
the: (a, b) bright type ($P Q > 0$, pulses); (c) dark type, (d)
gray type ($P Q < 0$, voids). Notice that the amplitude never
reaches zero in (d).

Figure 2.

The ratio $\eta = P/Q$ of the coefficients in the NLSE (\ref{NLSE}) is
depicted versus the wave number $k$ (normalized over $k_D$), for
several values of the dust parameter $\mu$;
in descending order (from top to bottom): 0.8, 0.9, 1.0, 1.1, 1.2.

\vskip 1cm

\newpage

% figure1

\newpage

% figure2


\newpage

\begin{thebibliography}{10}

\bibitem{PSbook} P. K. Shukla and A. A. Mamun, \textit{Introduction to Dusty
Plasma Physics} (Institute of Physics Publishing Ltd., Bristol, 2002).

\bibitem{Verheest} F. Verheest, \textit{Waves in Dusty Space Plasmas}
(Kluwer Academic Publishers, Dordrecht, 2001).

\bibitem{Krall} N. A. Krall and A. W. Trivelpiece,
\textit{Principles of plasma physics}, McGraw - Hill (New York,
1973); Th. Stix, \textit{Waves in Plasmas}, American Institute of
Physics (New York, 1992).

\bibitem{PSsolitons} For a review, see:
P. K. Shukla and A. A. Mamun, New J. Phys. \textbf{5}, 17.1 (2003).

\bibitem{IKPSDIAW} I.Kourakis and P. K. Shukla,
\textit{Physics of Plasmas} \textbf{10} (9), 3459 (2003);
\textit{Eur. Phys. J. D} \textbf{28}, 109 (2003).

\bibitem{Davidson1} R. C. Davidson and P. P. J. M. Schram,
\textit{Nuclear Fusion} \textbf{8}, 183 (1968).

\bibitem{Davidson2} R. C. Davidson, {\it{Methods in nonlinear plasma
theory}}, Academic Press (New York, 1972).

\bibitem{Infeld}
E. Infeld and G. Rowlands, Phys. Rev. Lett. {\textbf{58}} (1987).

\bibitem{Chakra1} N. Chakrabarti and M. S. Janaki,
{\it Phys. Lett.} A {\bf 305} 393 (2002).

\bibitem{Chakra2} N. Chakrabarti and M. S. Janaki,
{\it Phys. Plasmas} {\bf 10} 3043 (2003).

\bibitem{Sagdeev} R. Z. Sagdeev, in {\it Reviews of Plasma Physics}, Vol. 4.,
ed. M. A. Leontovich, Consultants Bureau (New York, 1966), p.52.

\bibitem{comment1} Eq. (\ref{L1}) was obtained from the
(Lagrangean) density equation, which is
reduced to: $\partial(n \alpha)/\partial \tau = 0$
by using the property (\ref{property1}); Eq. (\ref{L1})
then follows.

\bibitem{comment2} Eqs. (\ref{system-reduced}) are derived
from Eqs. (\ref{L1}, \ref{L2}, \ref{L3}), $E=-\nabla \phi$ and
(\ref{property1}), respectively. We have avoided the appearance of
$\alpha^{-1}$ -- cf. Eqs. (\ref{L4}, \ref{L5}) -- for analytical
convenience.

\bibitem{redpert} T. Taniuti and N. Yajima, \, J. Math. Phys. {\bf 10}, 1369
(1969); N. Asano,\, T. Taniuti and \, N. Yajima,
\, J. Math. Phys. {\bf 10}, 2020 (1969).

\bibitem{Hasegawa} A. Hasegawa,
\textit{Plasma Instabilities and Nonlinear Effects}
(Springer-Verlag, Berlin, 1975).

\bibitem{Fedele} R. Fedele, H. Schamel and P. K. Shukla,
{\it Phys. Scripta} T {\bf 98} 18 (2002); R. Fedele and H.
Schamel, {\it Eur. Phys. J. B} {\bf 27} 313 (2002);
Fedele, {\it Phys. Scripta} {\bf 65} 502 (2002).

\end{thebibliography}
\end{document}